\documentclass[sigconf]{acmart}
\AtBeginDocument{%
  }
\usepackage{multirow}
\usepackage{placeins}
\usepackage{hyperref}
\hypersetup{colorlinks,allcolors=black}
\usepackage{dblfloatfix}
\usepackage{enumitem}

\acmConference{Defective}{Amplified}{Contested}
\setcopyright{none}
\renewcommand\footnotetextcopyrightpermission[1]{}

 \usepackage{multirow}
 \usepackage{colortbl}   
 \usepackage{array}
 \usepackage{booktabs}

\newcommand{\nn}[1]{(\textit{n} = #1)}

\begin{document}

\title{Space and Time in Human Construals of AI}


\author{Jaime Banks}
\email{banks@syr.edu}
\orcid{0000-0002-7598-4337}
\affiliation{%
  \institution{School of Information Studies, Syracuse University}
  \city{Syracuse}
  \state{New York}
  \country{United States}
}

\renewcommand{\shortauthors}{Banks}

\begin{abstract}
Human–AI interaction is always situated—shaping and shaped by the space and time in which it is embedded. Research on interaction contexts focuses largely on what machines can be aware of; less is known about how non-experts, in everyday use, construe space and time as relevant to AI. Two exploratory online surveys of Large Language Model (LLM) users elicited open-ended accounts of space  (Study 1: \textit{N} = 75; spatial facets: proximity, composition, intersections) and of time (Study 2: \textit{N} = 85; temporal facets: speed, tempo, duration). Thematic analysis suggests humans variably locate AI in digital and physical spaces, in every space, and in no space at all. They also locate AI in present and future times, or as being outside of linear time altogether. Yet, in construals of AI, space and time are more than context. Interpreted through Schutz’s phenomenology of the lifeworld, themes suggest humans index AI egocentrically, relative to a situated “me-here-now,” through four dynamics: Position, reach, coupling, and reconfiguration. Those dynamics correspond with known relational domains among humans (consociates, contemporaries, and consociated contemporaries) but also point to additional domains: \textit{Achronic consociates} (sharing space but not time), \textit{pansociates} (saturating humans’ phenomenal space and time), and \textit{extrasociates} (located beyond human space and time, yet interactively accessible). Space and time thus function not only as the backdrop for human–AI interaction but as part of how users make (sometimes contradictory) sense of what AI is, what it affords, and how they relate to it.
\end{abstract}



\maketitle
\pagestyle{fancy}
\fancyhf{}                                   
\fancyhead[L]{\small Space and Time in Human Construals of AI}
\fancyhead[R]{\small Banks}
\fancyfoot[C]{\thepage}                      
\renewcommand{\headrulewidth}{0pt}       

\section{Introduction}
When interacting with conversational AI, like LLMs, the interface may appear to be configured the same, the timbre and cadence of response seem the same, and call-and-response operates about the same. However, the same question typed into the chat window during a 9 a.m. workplace meeting, a 6 p.m. barstool debate, or at 2 a.m. in bed during a bout of insomnia may manifest very different interaction experiences. The same system can be encountered across different spatial and temporal contexts, and those contexts may shape the subjective experience of a human-AI interaction (HAII). 

In human-machine interaction scholarship, context is often reduced to semantics, interaction goals, or sociotechnical structures embedding interactions. Although those context features are important, less attended to are the more fundamental considerations of contextualizing \textit{space and time} \cite{papneja2025self}. Spatial and temporal contexts are well theorized for physically embodied robots (e.g., proxemics; \cite{nahum2026advancing}) and machine actors in augmented reality (e.g., context-aware agents; \cite{wu2026conversational}), where spatial presence is overtly visible and designed and there are observable actions performed over linear time. Because LLMs are more lean in conveying such cues, the human experience of AI’s spatiotemporality may rely more on heuristics, intuitions, or folk logics. Those construals are not yet well understood. 

Toward addressing this gap, the present study charts the folk-theoretical territory of how humans think about space and time in relation to AI. Findings indicate time and space are relevant contextual factors (where and when it is), but they \textit{also} may also shape core construals of what an AI is, how it works, what it affords, and how they relate to AI.

\section{Theoretical Background}

Context is a setting for a particular situation, linking together situational elements and contributing to their meanings \cite{johns2018advances}. Interactions are always situated. In HAII, questions often turn to whether and how AI can be aware of the setting or situation; in that vein, context can be defined as the machine-engageable information relevant to the interaction, such as locations, states, and actors, and technologies \cite{dey2001conceptual}. Objective environments, however, are not necessarily contexts; they dynamically become contexts when an interactant identifies them as relevant to the interaction \cite{dourish2004what}. The interactant determining relevance may be a human interlocutor or a machine interlocutor participating in an exchange (see \cite{banks2020agent}).

To ground this investigation, I tentatively define HAII context as the setting(s) that give rise to and are constituted by communication between human and artificial agents, insofar as those agents perceive them to be relevant to that communication. This includes, but is not limited to, \textit{where and when} each agent is. Whether and how human users construe convergent or divergent spatial and temporal situatedness is an open question. It calls for  empirical attention because a situation bounded by time and space sets the parameters for what interactants can perceive of and respond to in one another \cite{goffman1964neglected}.

\subsection{Spatiality: A Brief Overview}
Space (noun) and spatiality (property) are contested concepts. Scholars distinguish spaces (locales with objective properties) from places (spaces made meaningful through experience; \cite{tuan1975place})—and both can manifest in and through the digital as well as the physical \cite{bowman2026egocentricity}. Space may be absolute and measurable, constituted by things’ relative positions, and/or constituted by relations and processes \cite{harvey2006space}. From these framings, I adopt an inclusive definition for this exploratory work: Space is a structural and semantic arrangement in which (and in relation to which) interaction unfolds such that actors, objects, meanings, and processes (and other concrete or abstract things) can be said to be “somewhere.” This definition supports an examination of three spatial properties of a situation, each of which can be actual or construed: proximity (relative distance), composition (the nature of things co-present in and constitutive of a situation), and intersections (the overlaps thereof). 

Current theory holds that electronic media (interfaces, and likely the agents they convey) shift how humans experience places because they reconfigure access—who can perceive and know what about whom \cite{meyrowitz1985no}. In effect, this means a “situation” can be defined by access to relevant information in an interaction, rather than merely by whether interactants are co-present. This reconfiguration can take multiple forms: Technologies can be present in a physical space (e.g., an AI is “in” a laptop or a server somewhere); they can be a space themselves (an app is a place humans go to interact with AI); they can hold internal representations of places (the model may know where the user is); and they can change the character of places (the office desk becomes a social space for encountering AI). There are additional possibilities for transcending space and place through simultaneity, distribution, hybridization, and re-presentation \cite{adams2011taxonomy}. There is mixed attention to human-situating physical space \cite{xu2026understanding}, digital spaces with shifted affordances and constraints (e.g., \cite{riva2025digital}), and hybrid spaces for convergent materialities (e.g., \cite{fettahoglu2026digital}). Other spatial configurations (as raised by \cite{adams2011taxonomy}) receive scant attention and most work takes space for granted.

\subsection{Temporality: A Brief Overview}
Time (noun) and temporality (property) are also treated in varied ways. Time may be organized by social activities and “fixed by the rhythm of collective life” (\cite{sorokin1937social}, p. 619), or it may be understood more generally as something enacted through individual or collective practice \cite{orlikowski2002about}. Time is regularized by patterns in order, length, temporal location (i.e., specific point in a timeline), and recurrence rates stitched together by social norms \cite{zerubavel1981temporal}. Some argue time is plural (laminating together cycles of biology, technology, and sociality) and that standardized “clock time” is separate from experience (e.g., \cite{adam2006time}). From these framings, I again adopt an inclusive definition to scaffold this study’s exploration: Time is the multifaceted arrangement of moments in which (and in relation to which) interaction unfolds, such that actors, objects, meanings, and processes (and other concrete or abstract things) can be said to be “somewhen.” This definition supports attention to three temporal facets of speed (slow to fast), tempo (pace and rhythm), and duration (how long something lasts). 

Time has received more scholarly attention than space, as a matter of process or context. The speed of an AI’s response and length of time to complete a task impact a range of human experiences (affective response, intentions, presence, trust) and interaction-product output \cite{brynjolfsson2025generative,gnewuch2022opposing}. Longitudinal studies point to relational decay over time \cite{skjuve2022longitudinal}, and other work finds that humans sometimes engage social AI functionally to help with the mundane routines of daily life \cite{depounti2026wild}. Less understood is how session duration impacts human experiences (often reported only descriptively or as a control). As with space, scholarship around time in HAII attends mostly to measured effects of specific variables or utility in particular domains, rather than how users make sense of time as a salient dynamic within interactions and as context for them.

\subsection{The Current Study}
As noted, there is a gap in our understanding of how space and time are relevant to human-AI interactions as seen through humans’ subjective experience and folk logics. To address this gap, I explore how people think and feel about how these contextual factors matter in human construals of AI and what that suggests about relevant theory and key questions. I ask: 

\textbf{RQ}: (How) do space and time figure into human construals of AI? 

This question is approached inductively, using the tentative definitions of space and time as a springboard for charting the territory. This descriptive work is an effort in signal detection—a systematic surveying of humans’ subjective construals for fruitful future directions.

\section{Method}
To address the research question, two online surveys were conducted: Study 1 on space and Study 2 on time. Each elicited human accounts of how important the factor and its facets (space: proximity, composition, intersections; time: speed, tempo, duration) are to how they think about AI. 

\subsection{Participants and Procedure}
I sampled 75 and 85 people from Prolific for Studies 1 and 2, respectively. Participation was limited to existing LLM users aged 18 years or older. Additionally, because space and time are subject to cultural differences in structure and norms and in how their relevance is linguistically indexed, for this exploration participants were limited to U.S. Americans fluent in English as a first language. In the 30-minute online surveys, participants answered questions about their AI use, the importance of space or time (and respective facets), and then responded to open-ended elicitations. Finally, they answered demographic and psychographic items. Prolific’s screening for bot-driven and AI-generated responses checked data provenance, and participants were paid through the platform. Data were collected in June 2026.

\subsection{Elicitations and Measures}
For each factor and facet, participants were given a high-level definition of the element (Table \ref{tab:definitions}), which was necessary to ensure non-expert comprehension of the question and was accepted as a trade-off for potential breadth of answers. Participants were then asked to rate “How important is [FACTOR/FACET] to how you think about AI?” on a 7-point Likert style scale (1 = not at all important to 7 = extremely important). Next, they were asked to explain why they answered as they did. The ratings were intended only as anchors for the open-ended response and are reported as descriptives. The primary data analyzed here were textual responses to open-ended elicitations. The factor (space or time) was presented first, and the facets were presented afterward in a random order.

\begin{table*}[htbp]
  \caption{Definitions of Space and Time factors and their facets, provided to participants.}
  \label{tab:definitions}
  \small
  \renewcommand{\arraystretch}{1.2}
  \begin{tabular}{@{}p{0.12\textwidth} p{0.44\textwidth} p{0.35\textwidth}@{}}
    \toprule
    \textbf{Factor/Facet} & \textbf{Provided Definition} & \textbf{Grounding conceptualizations for provided definitions} \\
    \midrule
    \textsc{Space} &
      ``the physical, digital, or symbolic locations of our experiences. We make sense of life by observing things in nearby spaces and imagining things in more faraway spaces.'' &
      \citet{trope2010construal}; \citet{tuan1975place} \\
    \hspace{1em}\textsc{Proximity} &
      ``how close or how far away things are'' &
      \citet{trope2010construal} \\
    \hspace{1em}\textsc{Composition} &
      ``the shape of spaces and what they are made of'' &
      \citet{desouzaesilva2006cyber}; \citet{harrison1996replacing}; \citet{harvey2006space} \\
    \hspace{1em}\textsc{Intersections} &
      ``the potential for spaces to overlap, cross, or join up'' &
      \citet{desouzaesilva2006cyber}; \citet{meyrowitz1985no} \\
    \midrule
    \textsc{Time} &
      ``a progression we experience as things happen one after the other. We make sense of life by remembering the before, living in this moment, and planning the next.'' &
      \citet{trope2010construal}; \citet{zerubavel1981temporal}; \citet{zimbardo1999putting} \\
    \hspace{1em}\textsc{Speed} &
      ``how fast or slow something is'' &
      \citet{gnewuch2022opposing}; \citet{wajcman2008life} \\
    \hspace{1em}\textsc{Tempo} &
      ``the pace and rhythm of something'' &
      \citet{adam2006time}; \citet{zerubavel1981temporal} \\
    \hspace{1em}\textsc{Duration} &
      ``how long something lasts'' &
      \citet{adam2006time}; \citet{zerubavel1981temporal} \\
    \bottomrule
  \end{tabular}
\end{table*}

Participant attributes were measured in two phases. At each survey’s start, participants indicated which LLMs they use at least weekly (multiple choice with optional entry), their weekly hours of interaction (open field), and LLM knowledge level (single item, 7-point Likert, 1 = total novice to 7 = total expert). At the end of the surveys, participants reported their age, country of residence, and how they experience space or time. For Study 1, the latter was a binary-response item to select whether they see space as controlling them or them controlling space (based on \cite{rotter1966generalized}). For Study 2, this was a multiple-choice item about the psychological experience of time \cite{zimbardo1999putting}, selecting a tendency to think about the past, live in the moment, or think about the future.

\subsection{Analytical Approach}
Open-ended responses were analyzed separately for each elicitation. Each set was subjected to inductive thematic analysis guided by Braun and Clarke’s \cite{braun2006thematic} approach, in six steps: Careful reading of the data, generation of initial (i.e., open) codes, iterative collapsing of codes into categories and themes based on similar concepts, review of themes against source data, theme definition/naming, and reporting. The coded units were the smallest identifiable complete thought, such that individual responses often had multiple codes. Initial codes were assigned based on manifest content, while successive rounds of collapsing were increasingly based on higher-order, interpreted relationships among the codes. Collapsing proceeded until themes were identified based on Braun and Clarke’s criteria of keyness (candidate theme utility in answering the research question) and prevalence (frequency in the corpus). Here, prevalence was benchmarked against an \textit{a priori}-established criterion of a theme accounting for at least 10\% of the coded units for a given elicitation. Coded units were not retained for theme consideration when they indicated space or time was irrelevant, commented on other topics, misunderstood the question, or rejected the premise; these units’ counts are reported. Results characterize how humans think about space or time \textit{when it matters to them}. Similar concepts could appear in multiple themes, depending on how they are phrased in responses. Positively and negatively valenced codes about the same topic are generally treated as two sides of the same conceptual coin, and so are integrated into nonvalenced themes. Themes are not harmonized across the elicitations or studies; they are treated as emerging independently from each elicitation and are, instead, synthesized in the discussion. 

My interpretive lens for this work is grounded in a material-semiotic ontology (things and their meanings are real), a relational epistemology (we come to know what is real by examining observable and inferred relationships among things and their meanings), and a pragmatic methodological leaning (it is prudent to match methods to the question, prediction, or problem posed). I am a regular user of general-purpose LLMs (primarily the Claude suite) and have moderate technical knowledge of them. I tend to (a) \textit{want} to feel control over spaces but actually feel controlled by them and (b) tend to think a lot about the future. This positionality necessarily shapes my interpretation of the data, along with my academic interests in human-machine communication, social cognition, and moral psychology and their respective bodies of theory.

\section{Study 1 Results: Spatiality in AI interactions}
Participants were, on average, 42.87 years old (\textit{SD} = 13.07, range 21-74). Most (86.7\%) indicated feeling they have control of spaces (versus spaces having control of them). They most reported using ChatGPT (81.3\%) and Gemini (69.3\%), followed by Claude, CoPilot, DeepSeek, Kimi, and Grok. They spend an average of 7.99 weekly hours (\textit{SD }= 8.77, range 1-40, Mdn = 5) using LLMs and indicated moderately high knowledge about LLMs (\textit{M} = 4.72, \textit{SD} = 1.06, range 1-7). Themes are presented in order of prevalence in the data, summarized in Table \ref{tab:space-themes} and explained below.

\newcolumntype{L}[1]{>{\raggedright\arraybackslash}p{#1}}

\begin{table*}[t]
  \caption{Thematic hierarchy for relevance of space and spatial dimensions to AI.}
  \label{tab:space-themes}
  \scriptsize
  \renewcommand{\arraystretch}{1.1}
  \begin{tabular}{|L{0.14\textwidth}|L{0.24\textwidth}|L{0.52\textwidth}|}
    \hline
    \textbf{Theme} & \textbf{Subthemes} & \textbf{Exemplar data excerpt} \\
    \hline
    \multicolumn{3}{|l|}{\textit{M} = 1.29 to 1.57 coded units per participant} \\
    \hline

    \rowcolor{gray!20}\multicolumn{3}{|l|}{\textbf{SPACE} \nn{118}} \\
    \hline
    \multirow{3}{=}{Spatial situatedness \nn{39}}
      & Functions in/through Digital Space \nn{11}
      & ``AI exists in digital spaces and affects how people interact, learn, and communicate'' \\ \cline{2-3}
      & Embedded in Physical Space (but maybe should not; \textit{n}\,=\,8)
      & ``I guess I do connect it with my desk/work area since I use it on my computer'' \\ \cline{2-3}
      & Situated Otherwise \nn{20}
      & ``I think of AI as just always there, in no one particular space.'' \\ \hline
    \multirow{4}{=}{Self-relative spatiality \nn{35}}
      & In my Space \nn{15}
      & ``it's like AI can sense where we are at when asking questions'' \\ \cline{2-3}
      & Space as Governance \nn{8}
      & ``I like the AI concentrated on my PC where I can choose how and when to use it.'' \\ \cline{2-3}
      & Governed by Space \nn{7}
      & ``where AI is used really affects how much I do trust it and also use it.'' \\ \cline{2-3}
      & Social Closeness \nn{5}
      & ``for me to feel close to it.... I need to have it in my close space. If it felt far away I would listen to it and I would be distant.'' \\ \hline
    \multirow{3}{=}{Liminality \nn{27}}
      & Bridges Space \nn{11}
      & ``helps people interact with [information] that may be near or far from them'' \\ \cline{2-3}
      & Impacts Real/Physical World \nn{11}
      & ``I think about the data centers and how destructive they can be to our space.'' \\ \cline{2-3}
      & Access to space \nn{5}
      & ``With AI, you are [able] to utilize space to whatever your needs may be.. From GPS to finding spaces, to AR videos where space is literally mapped out'' \\ \hline
    \multicolumn{2}{|L{\dimexpr0.38\textwidth+2\tabcolsep+\arrayrulewidth\relax}|}{Hardware dependence \nn{17}}
      & ``network servers are kept or arranged in a particular location or places'' \\ \hline

    \rowcolor{gray!20}\multicolumn{3}{|l|}{\textbf{PROXIMITY} \nn{106}} \\
    \hline
    \multirow{4}{=}{Proximity to Self \nn{44}}
      & Closeness to the Body \nn{17}
      & ``the proximity of the AI already is. I carry it with me.'' \\ \cline{2-3}
      & Close is Useful \nn{12}
      & ``it has to almost within hands reach to become a valuable asset'' \\ \cline{2-3}
      & Relevance to Life \nn{12}
      & ``AI feels more trustworthy when it directly affects my daily life'' \\ \cline{2-3}
      & Responds Uniquely to Me \nn{3}
      & ``I like it to be adjusted to me as a person... respond in a way that relates to me.'' \\ \hline
    \multirow{5}{=}{Proximity by Performance \nn{36}}
      & Bridges Me/World \nn{12}
      & ``fetching information from far and wide and bringing it to my laptop'' \\ \cline{2-3}
      & Has Impact \nn{10}
      & ``AI's impact depends on what is near or distant---information, people, decisions, and consequences.'' \\ \cline{2-3}
      & Function of Efficiency \nn{8}
      & ``milliseconds for a search to reach millions and billions of different areas'' \\ \cline{2-3}
      & Usability \nn{4}
      & ``Data proximity directly impacts latency in our small world'' \\ \cline{2-3}
      & Future Factors \nn{2}
      & ``Until [A.I.] is given robotic bodies, humans can still have some solitude.'' \\ \hline
    \multirow{5}{=}{Relational Proximity \nn{26}}
      & Interpersonal Closeness \nn{10}
      & ``if you get wrapped up in it, you feel close to it physically and mentally'' \\ \cline{2-3}
      & Need/Dependence \nn{4}
      & ``I have come to depend on AI, at least to an extent, and I feel reassured by the idea of being close to it.'' \\ \cline{2-3}
      & Depends on task/context \nn{4}
      & ``if I have to sit down at a computer to use it, it stops feeling like a helpful assistant and just feels like a stiff, formal tool.'' \\ \cline{2-3}
      & Intensity \nn{4}
      & ``the closeness is mostly about how much you interact with it'' \\ \cline{2-3}
      & Ontological Considerations \nn{4}
      & ``AI could be considered intellectually and emotionally close to humans'' \\ \hline

    \rowcolor{gray!20}\multicolumn{3}{|l|}{\textbf{COMPOSITION} \nn{94}} \\
    \hline
    \multirow{5}{=}{Situatedness \nn{28}}
      & Physically Situated \nn{14}
      & ``this tool is on a server somewhere'' \\ \cline{2-3}
      & Access to Spaces \nn{6}
      & ``depending on what you are looking to accomplish it is important that it is connected to all things that are [relevant]'' \\ \cline{2-3}
      & Unbound/Unconstrained \nn{5}
      & ``The digital realm is nearly infinite'' \\ \cline{2-3}
      & Digitally Situated \nn{2}
      & ``It operates within spaces of information.'' \\ \cline{2-3}
      & Hostility of Human Space \nn{1}
      & ``the nature of a social space... may be more hostile to ai'' \\ \hline
    \multirow{3}{=}{Operation Shaping \nn{25}}
      & Comp. Shapes Outputs \nn{10}
      & ``Clean, structured datasets and robust network infrastructure are vital for accurate model outputs.'' \\ \cline{2-3}
      & Comp. Shapes Interaction \nn{8}
      & ``composition of the Ai space itself... makes it easy to use Ai'' \\ \cline{2-3}
      & Interface Composition \nn{7}
      & ``if the layout is bad, using the tech just feels clunky and annoying'' \\ \hline
    \multirow{3}{=}{Materiality \nn{21}}
      & Is (in) Soft \nn{10}
      & ``Its just computer code'' \\ \cline{2-3}
      & Is Without Composition \nn{7}
      & ``It is like an idea, it is there but does not take up space.'' \\ \cline{2-3}
      & Is Physical \nn{4}
      & ``it feels physical in the way a computer does'' \\ \hline
    \multirow{2}{=}{Outputs or Effects \nn{20}}
      & Textual Composition \nn{11}
      & ``Questions are asked and answered in written or maybe spoken word that can have composition...ie.structure'' \\ \cline{2-3}
      & Environmental Composition as Side Effect \nn{9}
      & ``AI impacts the space we operate in via innovation in technology and physical applications of ideas.'' \\ \hline

    \rowcolor{gray!20}\multicolumn{3}{|l|}{\textbf{INTERSECTIONS} \nn{91}} \\
    \hline
    \multirow{3}{=}{Material Intersections \nn{57}}
      & Soft-Space Intersections \nn{24}
      & ``LLMs crossing over various sources of information to acquire the answers needed to inform users'' \\ \cline{2-3}
      & Hard-Space Effects \nn{18}
      & ``the spaces of AI overlapping with our daily lives can influence our lives'' \\ \cline{2-3}
      & Liminal-Space Intersections \nn{15}
      & ``I think it's important for all of us to learn how to interact with intersect with AI and evolve with it in our world.'' \\ \hline
    \multicolumn{2}{|L{\dimexpr0.38\textwidth+2\tabcolsep+\arrayrulewidth\relax}|}{Egocentric Intersections \nn{13}}
      & ``need for my AI to be able to connect past conversations... like my AI knows me'' \\ \hline
    \multicolumn{2}{|L{\dimexpr0.38\textwidth+2\tabcolsep+\arrayrulewidth\relax}|}{Intersectional Ontology \nn{11}}
      & ``AI itself is fundamentally built from relationships rather than isolated facts'' \\ \hline
    \multicolumn{2}{|L{\dimexpr0.38\textwidth+2\tabcolsep+\arrayrulewidth\relax}|}{Systemic Intersections \nn{10}}
      & ``We intersect AI in art, music, customer service, social media, the news, and basically everywhere.'' \\ \hline
  \end{tabular}

  \par\smallskip
  \noindent{\scriptsize\textit{Note:} Words in brackets are author-edited for clarity.}
\end{table*}

\subsection{SPACE}
On average, respondents were equivocal or ambivalent about the relevance of space to their thoughts about AI (\textit{M} = 3.87, \textit{SD} = 1.99). Four themes were induced for participants’ considerations of “space,” in general: 

SPATIAL SITUATEDNESS: Participants discussed spaces where AI exists or is situated. Very few instances overtly noted that AI exists in physical space; where they did, AI was suggested to relate to specific work areas, with digital spaces existing within physical spaces, or with AI exploring terrestrial environments for answers. Some also said AI \textit{should not} have or does not need physical spaces. More commonly, AI was noted to function within or through digital spaces—it exists in such environments, including cloud or development environments, online milieu, or spaces made of vectors and equations. AI also intrudes upon digital domains of human work. But most commonly, participants characterized AI as being otherwise situated—it is nowhere, it doesn’t take up space. It is omnipresent and omnitemporal, or because it can be called up it exists in \textit{all} spaces. It occupies spaces we do not and moves beyond what we can experience. Because of this, its imagination is not limited and it does not suffer the restrictions of earthly occupation. For some, AI \textit{is} or \textit{is like} a space—a library, a locale for communication, an environment to think clearly within.

SELF-RELATIVE SPATIALITY: The relevance of space to AI was frequently discussed in the ways AI was relevant to the self—one’s perceived location, joint situatedness, and forms of agency or emotions. AI’s space was seen in relation to one’s own space—it is wherever I am, it is here, it is when and where I call it. Participants saw this co-presence both positively and negatively. Some saw it as vital to AI’s functioning, as it must be “here” in order to understand and know the person’s world and to function as a cognitive extension. Others saw it as part of an improved mental space, and others still saw it as a surveillance device—always judging, always listening. Some described social closeness and presence, feeling close to AI as a mental or emotional (and not physical) exercise, getting lost in interactions, and finding the “symbolic” proximity reassuring. Two contrasting views of governance also emerged: Space as a form of governance over interactions (setting the purpose, norms, expectations, mindsets) versus space as a way to govern their exposure to AI (tethering AI to themselves via a phone or anchoring it to a PC so they can walk away from it). 

LIMINALITY: Adjacent to AI’s situation in digital, physical, or other spaces are the capacities for AI to bridge, aggregate, or otherwise sit at spatial intersections. Often, this liminality was related to the organization of information across and within digital and physical spaces, as AI consolidates information in a conversation as if the human was talking to a multitude of people. That coordination helps to connect information that is near and far and is executed through the marshalling of separate processes in different spaces. This in part comes from AI’s ability to access information \textit{about} spaces—from GPS to augmented reality data—such that spaces are things for AI to study and imagine, and for AI to help us study and imagine. Liminality was just as often about how AI functioning can have impact on physical-space phenomena, from environments and society to individual psychology. One participant mused that it is possible that our interactions with AI, in its digitally tethered liminality, could eventually alter human relationships with physical places. 

HARDWARE DEPENDENCE: Although data included very few mentions of AI existing in physical space, there were many instances noting that AI was not in physical space but relied on devices or infrastructure in physical space. Most of these referenced data centers: They take up space, consume physical resources, and are sometimes seen as intrusive in or destructive of spaces they occupy. AI are also seen as tethered to or deployed through specific devices, though that is less about spatiality and more about one’s relationship to it.

Aside from these themes, \textit{n} = 30 units indicated space is irrelevant or inapplicable to AI, \textit{n} = 10 offered extraneous commentary, and \textit{n} = 2 challenged or struggled with the question or premise.

\subsection{PROXIMITY}
Ratings for spatial proximity’s relevance to AI were, on average, just above the scale midpoint (\textit{M} = 4.15, \textit{SD} = 2.02). Three themes were induced for participants’ considerations of “proximity” as closeness or farness:

PROXIMITY TO SELF: Notions of spatial proximity were most considered through an egocentric lens—the ways in which AI is proximal or distal to oneself, broadly construed. Participants considered the closeness of AI to their physical bodies as it is carried, embedded in nearby objects, or confined to or accessed through devices (e.g., phones, computers, keyboards). Bodily closeness shapes AI’s utility and access, as AI must actually be close or feel as though it is close to be useful, though some preferred greater distance. Self-proximity extended to AI’s relevance to daily life and activities, including utility in work/home tasks or in decision making, and as part of everyday routines (especially embeddedness in personal phones). In these ways, proximity shapes evaluations, where close AI are more trustworthy and closeness prompts considerations of privacy and personal choice. Some suggested that closeness manifests when AI feels that it is tailored uniquely to them, relating to and responding to their needs. 

PROXIMITY BY PERFORMANCE: Proximity can be interpreted based on an AI’s operational quality, capacities, or outcomes. Speed and accuracy (seen as a function of compute, data pipeline, and latency) generate felt proximity. So, too, does usability (experiencing flow, naturalism, peak performance, and reliability). Performance impacts consider how operations and efficiencies occur in or impact physical spaces: AI’s grounding in data centers and influence on the environment makes it feel close. This construed proximity can affect user considerations when closeness makes salient the costs and benefits and when distance prompts considerations of more general ethics. AI capacities also afford intersections of speed, accuracy, and physical links to bridge gaps between humans and the world by bringing information from “everywhere” to the user, helping to find information about near or distant spaces, and making the inaccessible more real. Participants suggested that, by quickly synthesizing information, AI narrows the distance among spaces and between spaces and themselves. Future factors (AI growth and embodiment) may change the dynamics of how designed AI performance relates to proximity.

RELATIONAL PROXIMITY: Relational proximity is a matter of interpersonal, interactive, and ontological closeness between human and AI. Interpersonal facets involve affective and social connectivity: Desires for AI to be personable, act as a partner, effect social realism, build bonds, and feel mentally and emotionally close; conversely, some rejected the notion of emotional closeness and intimate knowing. Some suggested proximity depends on the intensity and outcomes of interactions, so the interactions influence the very functioning of AI. Others escalated closeness by interpreting proximity as need or dependence that may foster discomfort or reassurance. This relationality is complicated by ontological considerations arising when AI is compared to humans as rival intellects, as inhuman, as unnatural, as less-than-tactile. Practical interactions maintain instrumental distance while personal interactions invite relational proximity (and sometimes the inverse, with proximity impacting intuitive forms of interaction).

Among coded units, \textit{n} = 9 units account for expressions that spatial proximity is irrelevant or inapplicable to AI, \textit{n} = 4 offered extraneous or ambiguous commentary, and \textit{n} = 1 challenged or struggled with the question or premise.

\subsection{COMPOSITION}
Importance of spatial composition to AI was again moderate (\textit{M} = 3.76, \textit{SD} = 1.95). Four themes were induced for participants’ considerations of “composition” as the shape of space and its make-up:

SITUATEDNESS: Most common were mentions of the composition of spaces in which AI is situated or to which it is linked. Most frequently participants said AI is situated in physical space, usually by being contained in, supported by, or used through physical devices. Some said AI is unconstrained by space of any composition—it exists anywhere in any space, outside of limited space, or in digital space seen as infinite. Few said AI exists in digital “realms” or information spaces, or sometimes in hostile human social spaces. Sometimes, AI was seen as affording humans access to the digital (e.g., simplifying access to vast information, allowing mediated entry to systems), though at the direction of controlling humans or regulation.

OPERATION SHAPING: Spatial composition was seen as shaping how AI operates in interactions. In one vein, context composition shapes behaviors or outputs. Notably, this claim was most often rather abstract, referring in general terms to shaping meaning, shaping what it can and cannot do, and determining how it works. Two of those suggestions imbued the AI with a degree of subjectivity or at least reactivity, suggesting AI responds to how a space is structured. Context composition also shapes interaction dynamics, as different environments require different types of inputs (e.g., a busy home requires voice inputs while a quiet office permits more concentrated typing), create the conditions for ease of use and trust, and foster perceptions that the machine is functioning in one’s own space and so subject to those expectations. Some interpreted spatial composition in terms of interface layout and functions, characterizing it (via layouts, arrangements, set-up, accessibility, simplicity) as impacting AI operations—and so the experience of interactions. 

MATERIALITY: Participants considered the material composition of AI itself. Most suggested AI is absent any composition because code, software, program, and the digital does not take up space or because it is like a void, an abstraction, or an idea). Some said AI is physical, composed of physical machines or having physicality like machines.

OUTPUTS OR EFFECTS: Spatial composition was also seen in terms of AI outputs or effects. Most often this referred to texts AI produces, which vary in compositional qualities—replies that are more vague or specific, more or less comprehensive and structured, of varying formats, and perhaps messy or verbose. Nearly as often, participants discussed the composition of environments as a side-effect or impact of AI operation, though one person suggested that AI could actually compose space as a kind of task. Through that lens, AI contributes to the composition of the world by being housed in physical environments requiring uses, interacting with people as interactants, altering human work environments, and translating ideas from abstractions to applications. 

Among coded units, \textit{n} = 15 units account for expressions that spatial composition is irrelevant or inapplicable to AI, \textit{n} = 11 offered extraneous or ambiguous commentary, and \textit{n} = 1 challenged or struggled with the question or premise.

\subsection{INTERSECTIONS}
Spatial intersection was the facet deemed most relevant to AI (\textit{M} = 4.71, \textit{SD} = 1.70). Four themes were induced for participants’ considerations of “intersections” as the ways spaces may overlap or adjoin:

MATERIAL INTERSECTIONS: Spatial intersections focused on materiality constitute the majority of coded units; these units note the composition of space in which the considered elements converge. Soft-space intersections concern AI’s role as an actor or intermediary within spaces of information, logics, software, or intangible or technical system elements. AI interacts with, combines, or crosses paths with other software/systems (engines, apps, software, systems, and processes). It convenes and synthesizes information across multiple application domains and sources and may intersect with other artificial agents—working together or even merging with them. Hard-space effects concern the intersection of AI’s direct or indirect material consequences: Shifts in productivity, altered human-human/tech relations, making personal spaces conspicuous, job displacement, and reconfigured social contact. Liminal-space intersections emphasize the meeting of digital and physical through interfaces, interactivity spaces, and hybrid social spaces. These are places where physical entities can access digital information, where past chat can inform present, where human and machine can be co-present—chat, blend, co-exist. 

EGOCENTRIC INTERSECTIONS: Some spatial intersections were considered through an egocentric lens, as participants detailed the way AI may be said to intersect adaptively across “my” spaces: Facilitating transitions from work to home, appearing where one needs it, bringing information to a situated person. 

INTERSECTIONAL ONTOLOGY: AI was seen by some as an intersectional site itself. It was described as comprising spaces where cognitive capacities, neural architectures, and functional domains converge. It was sometimes seen as existing in a multiplex plane of its own—persistent, omnipresent, and connecting a range of otherwise disconnected spaces.  

SYSTEMIC INTERSECTIONS: AI may constitute spaces occupied by complex systems, which it assembles by bringing together heterogeneous elements (people, platforms, contexts) and simultaneously creating and being embedded within economies, societies, and institutions. Some systems are ideological assemblages, manifesting when AI aggregates (even contests) information and knowledge. 

Among coded units, one indicated that spatial intersections are irrelevant or inapplicable to AI, \textit{n} = 9 offered extraneous or ambiguous commentary, and \textit{n} = 1 challenged or struggled with the question or premise.

\section{Study 2 Results: Temporality in AI Experiences}
On average, participants were 38.01 years old (\textit{SD} = 10.56, range 18-65). They were split in their temporal orientations: Most said either they live in the moment (41.2\%) or think about the future (40.0\%), while fewer said they think about the past (18.8\%). They most reported using ChatGPT (78.8\%) and Gemini (63.5\%), followed by Claude, CoPilot, and DeepSeek. They spend an average of 7.58 weekly hours (\textit{SD} = 9.18, range 0.5-55, Mdn = 5) using LLMs and indicated a moderately high level of knowledge about LLMs (\textit{M} = 4.72, \textit{SD} = 1.08, range 1-7)—comparable to Study 1. Themes for each elicitation are presented in order of prevalence in the data, summarized in Table \ref{tab:time-themes} and explained below. 

\begin{table*}[t]
  \caption{Thematic hierarchy for relevance of time and temporal dimensions to AI.}
  \label{tab:time-themes}
  \scriptsize
  \renewcommand{\arraystretch}{1.1}
  \begin{tabular}{|L{0.14\textwidth}|L{0.24\textwidth}|L{0.52\textwidth}|}
    \hline
    \textbf{Theme} & \textbf{Subthemes} & \textbf{Exemplar data excerpt} \\
    \hline
    \multicolumn{3}{|l|}{\textit{M} = 1.32 to 1.86 coded units per participant} \\
    \hline

    \rowcolor{gray!20}\multicolumn{3}{|l|}{\textbf{TIME} \nn{151}} \\
    \hline
    \multirow{4}{=}{Time savings \nn{83}}
      & Saves time \nn{36}
      & ``when it works correctly, it saves time'' \\ \cline{2-3}
      & Work optimization \nn{33}
      & ``help [complete] tasks so much faster than if I... do it on my own'' \\ \cline{2-3}
      & Fast by design/function \nn{7}
      & ``can produce summarizations of larger articles'' \\ \cline{2-3}
      & Balance speed and accuracy \nn{7}
      & ``I would not mind if it took more time if I knew it was giving me accurate information.'' \\ \hline
    \multirow{2}{=}{Temporal situatedness \nn{35}}
      & Situated in time \nn{28}
      & ``AI doesn't exist in vacuum and relies entirely on a progression from past data to future impact'' \\ \cline{2-3}
      & Exists separately from time \nn{7}
      & ``AI to me is stagnant and time does not affect it in the same way it affects a human'' \\ \hline
    \multirow{4}{=}{Influence on human experience of time \nn{33}}
      & Moderates relationship with time \nn{13}
      & ``there are times to where AI Can cause people to lose track of time'' \\ \cline{2-3}
      & Entanglements with understanding/remembering personal history \nn{10}
      & ``it help us remember or recall things from the past that we can learn from'' \\ \cline{2-3}
      & Prospective change \nn{6}
      & ``time is a constant that confirms fate, with time , we get to know the advancements of AI .'' \\ \cline{2-3}
      & Retrospective change \nn{4}
      & ``I do miss certain moments where AI wasn't the core topic of discussion.'' \\ \hline

    \rowcolor{gray!20}\multicolumn{3}{|l|}{\textbf{SPEED} \nn{156}} \\
    \hline
    \multirow{4}{=}{Egocentric stance \nn{59}}
      & Preference for speed \nn{29}
      & ``I want quick answers and responses'' \\ \cline{2-3}
      & Speed relative to self-speed \nn{19}
      & ``I need the AI to be at my speed at all times and that means keeping up with me always'' \\ \cline{2-3}
      & Speed of conversation \nn{7}
      & ``I want to keep a quick and fluid conversation. If it takes too long then it disrupts my flow'' \\ \cline{2-3}
      & Speed requires commitment \nn{4}
      & ``make things speedier... [but] discerning information takes human intervention.'' \\ \hline
    \multirow{3}{=}{Speed (+ truth) as ideal \nn{57}}
      & Balance speed and truth \nn{21}
      & ``Care more about right answers than fast ones. Slow is fine for my accuracy and interpretation.'' \\ \cline{2-3}
      & Speed is the point \nn{19}
      & ``If an AI works slow there is no reason to use it'' \\ \cline{2-3}
      & Speed is utility \nn{17}
      & ``The faster it responds the more useful it is in real situations.'' \\ \hline
    \multirow{2}{=}{Utility impacts \nn{20}}
      & Impact on work efficiencies \nn{16}
      & ``... impressive speeds. This increases productivity and workflow.'' \\ \cline{2-3}
      & In-the-moment usefulness \nn{4}
      & ``responding to real life situations at the moment it is calling for'' \\ \hline
    \multirow{2}{=}{Contextual factors \nn{20}}
      & Speed of context \nn{14}
      & ``it is used in a wide variety of settings, some of which carry time constraints'' \\ \cline{2-3}
      & Technical factors \nn{6}
      & ``it's also dependent on my own internet speed'' \\ \hline

    \rowcolor{gray!20}\multicolumn{3}{|l|}{\textbf{TEMPO} \nn{126}} \\
    \hline
    \multirow{3}{=}{Attunement \nn{60}}
      & Cadence disruption \nn{28}
      & ``i need it to work at my pace because if it [doesn't], it will affect my work and disrupt the flow.'' \\ \cline{2-3}
      & Social/natural cadence \nn{23}
      & ``A positive interaction with AI is usually filled with back and forth messages that need to mesh well.'' \\ \cline{2-3}
      & Flow states \nn{9}
      & ``It should vibe with you when interacting with it.'' \\ \hline
    \multirow{3}{=}{Adaptivity \nn{35}}
      & Adjust to match me \nn{17}
      & ``I like when AI can keep up with my, very often, stream of consciousness thought patterns'' \\ \cline{2-3}
      & Adjust to an even/steady pace \nn{10}
      & ``builds a rhythm to the conversation... natural and steady.'' \\ \cline{2-3}
      & Adjust to the purpose/task \nn{8}
      & ``good AI adapts its pace, slow for learning , first for emergencies and rigid tempo feels robotic'' \\ \hline
    \multirow{3}{=}{Operational dynamics \nn{31}}
      & Just go fast \nn{21}
      & ``I want the information I asked for to be delivered quickly'' \\ \cline{2-3}
      & Diagnostic value \nn{7}
      & ``shows a level of competency... I need it to be in rhythm. When it's not, it's like a car battery not connecting all the way.'' \\ \cline{2-3}
      & Collaboration synergies \nn{3}
      & ``I like the pace of Claude because it shows little lines of text to simulate what it is doing'' \\ \hline

    \rowcolor{gray!20}\multicolumn{3}{|l|}{\textbf{DURATION} \nn{103}} \\
    \hline
    \multirow{2}{=}{Value orientations \nn{64}}
      & Valued efficiencies orientation \nn{35}
      & ``duration between putting in prompts and getting responses is huge... there's no need to use AI.'' \\ \cline{2-3}
      & Productive lag orientation \nn{29}
      & ``be [thorough] and ask lots of questions instead of getting the bare minimum. I want to explore things I'm missing or haven't thought of'' \\ \hline
    \multirow{2}{=}{Interaction dynamics \nn{20}}
      & User v.\ AI agency/governance \nn{14}
      & ``we ultimately control how long our interaction with AI lasts. It is available as we need, not the other way around.'' \\ \cline{2-3}
      & Exchange is depth \nn{6}
      & ``by having deeper conversations with AI it will be able to learn how to understand and communicate better'' \\ \hline
    \multirow{2}{=}{Duration as capacity \nn{19}}
      & Endurance \nn{13}
      & ``can support me for as long as i need... during longer tasks or conversations'' \\ \cline{2-3}
      & Technical/monetization constraints \nn{6}
      & ``the conversations get cut short [on free accounts]... i am stonewalled into waiting four or five hours to chat again.'' \\ \hline
  \end{tabular}

  \par\smallskip
  \noindent{\scriptsize\textit{Note:} Words in brackets are author-edited for clarity.}
\end{table*}

\subsection{TIME}
Participants leaned toward time, in general, as relevant to how they think about AI (\textit{M} = 5.07, \textit{SD} = 1.49). Three induced themes characterize participants’ considerations of time:

TIME SAVINGS: This highly instrumental theme focuses on AI’s potential for speed and efficiency, often compared to humans. AI is seen as fast by design in its core operations (search, generation, summarization). This speed frames AI as a time-saving tool that is particularly valued in fast-paced or high-stakes contexts (e.g., fast-paced or stressed environments, time-driven tasks). Some note the downside of this speed in the form of distraction for core tasks or rushed errors; others suggest the value of speed only alongside accuracy since errors are costly. Underlying this theme is a logic of work optimization (i.e., time is money) in which people value efficiency gains for ease, offloading undesired tasks, or reallocating time to other valued tasks, activities, or relationships. Participants are willing to put in their own time to learn tools for optimization and value reliable functionality over the long term; there are also high expectations that an AI must not miss moments in real-time tasks and must not trade speed for correctness. 

TEMPORAL SITUATEDNESS: This theme comprises competing understandings of whether AI has a relationship to time at all. Most commonly, participants see AI as situated in linear time either as a matter of course (as are all things in the world) or as inherent to its functioning: AI is dependent on past and current information to reason or to make projections, requiring up-to-date information. AI itself is subject to time’s effects, interpreted as rapid technological development (e.g., through learning, training, practice) while also being subject to bugs and producing negative effects on humans and the environment. Others see AI as existing outside of time—it is stagnant, unaffected, with no past or future. Unaligned with subjective temporal experiences (sometimes articulated as a human construct rather than an objective reality), it cannot understand that time has passed and cannot see things through a temporal lens.

INFLUENCES ON HUMAN EXPERIENCE OF TIME: The focus in this theme shifts from efficiencies to how humans subjectively experience the effects and nature of time. AI can influence how one spends time and derives meaning from its passage (e.g., mitigating time spent ruminating in indecision, supporting a focus on or planning for the future), can help one to live in the moment or catalyze the interaction window as a meaningful moment in itself (even to the point of getting lost in the moment), and can support one’s connection with their own life trajectory by helping one to remember or to predict. Sometimes this extends to facilitating connections with the past more generally (e.g., relevant histories, creating and saving memories) with an acknowledgment that human memories degrade. Inversely, some suggest their personal histories contain elements that shape how they view AI, having grown up with rapid technological change and having seen their own use patterns shift over time. A smaller subset of responses note how time uncovers or generates change. Some participants refer to AI’s cultural impact over time, expressing nostalgia for times when AI was not the focus and when there was not so much uncertainty—nothing is as it was before. Others suggest time will bring insight and answers to questions around AI’s impacts on human life and the progression of AI’s capabilities.

Among coded units, \textit{n} = 10 units account for expressions that time is irrelevant or inapplicable to AI, \textit{n} = 1 offered extraneous commentary, and \textit{n} = 1 challenged or struggled with the question or premise.

\subsection{SPEED}
Speed was highly relevant to how respondents think about AI (\textit{M} = 5.75, \textit{SD} = 1.19). Four themes were induced from considerations of speed as the fastness or slowness of a thing:

EGOCENTRIC STANCE: This theme frames the question of AI and time as a matter of self-relevance. Participants generally preferred and expected immediacy and directness, and many were frustrated by or simply disliked slow responses; a mild accommodation came in accepting longer response times so long as they were reasonable. The AI must have an appropriate speed relative to the self—participants acknowledge its speed but expect it to keep up with or be faster than the human and to keep pace with the speed of typical mediated human conversation; otherwise, many indicated, they would disengage to do tasks themselves or find another AI. For conversation in particular, lag is seen as detracting from the experience, as people want quick and fluid conversations that are also not too fast (lest they feel overwhelming or impersonal). Some recognize their own contributions to this figuration—they themselves impact speed through input quality and tempo as well as accuracy checks.

SPEED + TRUTH AS IDEAL: This theme addresses a tension between speed as a means (balanced against other valued qualities) and speed as an end in itself (even at the expense of those qualities). Participants broadly saw a combination of speed and correctness as the ideal and preferred AI performance—even so far as to say that this is AI’s \textit{promise}. For many, speed is inherently the point and lag or slowness defeats the purpose—if it’s not fast it’s moot and with the advantage erased, many would rather do the work themselves. For others, speed is the benefit but it cannot cancel out requirements for accuracy, reliability, or correctness; participants report willingness to wait longer for checked, thoughtful, expert-level responses over fast but false ones. Alongside the ends/means friction are tandem heuristics in which users read speed as a signal of other attributes: Fast responses imply confidence, competence, smartness, utility \textit{but also} may imply potential error, slop, and more general sacrificed quality. 

UTILITY IMPACTS: This pattern links the pace of exchange to productive outcomes, as speed is seen as directly impacting work efficiencies—either increasing or decreasing. Response speed can impact thought and workflows by optimizing or disrupting them, cutting the time, load, or cost of jobs, and sometimes maximizing human performance. Outside of formal work, AI response speed drives utility in everyday life as people use generated information in conversations, solve realtime problems, deal with situated dilemmas, and meet whatever the moment calls for. If AI is too slow to serve this purpose, people will move on. 

CONTEXTUAL FACTORS: Contextual considerations are qualifiers to the above ideals, orientations, and utilities. Participants cite dynamics of social life that drive speed requirements: The speed of contemporary life means people are busy, time is money, and inefficiencies mean you fall behind. Others call for contextually appropriate speed—AI responses on par with the complexity of the request or the job, or that moderate to meet the needs of time constraints versus use in leisure time. Some noted that the context drives AI interaction in the first place—they wouldn’t use it if they weren’t in a rush. Others still considered technical factors as moderating expectations, as internet speeds, input/output complexity, and model differences may shift performance. In these ways a “good” speed is contingent on the situation in which the interaction operates. 

All responses to this elicitation were coherent and agreeable, and none were deprecated for stating speed is irrelevant or difficult to grasp. 

\subsection{TEMPO}
Tempo was above the scale midpoint in importance (\textit{M} = 4.79, \textit{SD} = 1.62). Three themes were induced from participants’ comments on tempo as pacing and rhythm:

ATTUNEMENT: Attunement captures the subjective, experienctial character of interaction cadence. Naturalness qualifies “good” interaction tempos as those that are conversational, flowing, and calibrated to the topic. They were described as smooth rather than robotic and as humanlike in their ease and reciprocity. When present, those qualities make interaction more comfortable and enjoyable while violations make it feel forced. In part from that naturalness comes the achievement and maintenance of flow states described as being in-sync or vibing, while disruption breaks that vibe. The results of disrupted attunement were commonly discussed. When an AI’s tempo is misaligned with human user preferences or abilities, it disrupts focus, causes frustration (likened to a bad coworker), and risks disengagement. In particular, AI delivering too much information too quickly (especially for novel topics) is overwhelming, though some simply enact their agency by reading at their own pace without concern for the AI’s pace.

ADAPTIVITY: Participants frequently detailed an expectation that AI adapt its tempo to the moment. Adaptation is most often described as matching the pace of the human interlocutor in their behaviors, skill levels, and preferences, and then adjusting to changes therein. More generally, participants noted a desire for AI to keep up with the speed of life and of streams of consciousness. Some acknowledge the interaction cadence depends on their reading speed and prompting approach. Others call for adaptation to the purpose or task in the interaction—the pace of information requirements, the complexity, and whether the exchange is focused on learning, reflection, or perhaps an emergency. Tempo matters more when outcomes are time-sensitive. Embedded in these preferences are tensions in the value of consistency, as some suggest evenness and steadiness is useful while others suggest a good tempo can vary based on complexity and quality requirements.

OPERATIONAL DYNAMICS: This theme comprises the mechanical and signaling functions of tempo, largely separate from their felt experience. Most commonly, respondents acknowledged that AI operates at a fast pace (with quick successive call-and-response behaviors) though fast tempos are effectively invisible when consistently achieved; this description is paired with a desire for AI to \textit{just go fast} in giving quick, direct, efficient answers—because waiting is wasting time. Interaction tempo also serves a diagnostic function as it signals (un)reliability or trustworthiness as well as processing competencies and thinking operations. Pacing influences collaboration by dictating the cadence of conversation while also fostering positive or negative outcomes. For instance, tempo can create productive energy and manage burnout, or a comfortable pace can aid in understanding and retention while rapid-fire delivery undermines it. 

\textit{N} = 21 units expressed tempo is irrelevant or inapplicable to AI, \textit{n} = 6 offered extraneous commentary, and \textit{n} = 4 challenged or struggled with the question or premise.

\subsection{DURATION}
Duration leaned high in its importance to participants (\textit{M} = 5.16, \textit{SD} = 1.68). Three themes were induced from participants’ considerations of duration relative to AI, framed as how long something lasts:

VALUE ORIENTATIONS: Participants varied their opinions of how time spent in interactions has value or not, with two higher-order classes. The first is based on valued efficiencies. This cluster of sentiments favored brevity and economical session durations—don’t dwell on topics, don’t dawdle with language (e.g., pleasantries), and offer only what’s necessary. A more lenient version of this orientation called for balancing session duration with quality—be helpful not chatty, give enough but don’t drag, lean toward fast-but-correct, and do not lag without adding value. Without these efficiencies, time invested in longer sessions loses value and stresses fragile users and resource systems. The second can be characterized as a productive-lag orientation. This orientation accepts and even welcomes longer sessions as a matter of practicality or preference. In this orientation, interactions can go on as long as needed to get an answer or set the conditions for \textit{answering}. This acceptance is often a function of situational factors (e.g., external deadlines or time frames, particular tasks or goals, unfolding needs for follow-up questions). Some simply want more detail, want opportunities to explore, need to sit with ideas to understand them, simply enjoying dialogue, or prefer deep dives or social exchanges that take longer.

INTERACTION DYNAMICS: Participants also said duration is tied up in how AI agents and their exchange qualities unfold. Participants often overtly contended that the human user governs duration. A session goes on as long as they want: They lead it, control it, determine follow-up questions. Generally, they felt no need to accommodate AI when it prompts them for response and simply stop talking when they are finished. Others more implicitly accounted for AI’s agentic influence  on duration, alluding to the point at which AI thinks it is finished and favoring its skill in knowing when to \textit{stop}. Some framed prompting as asking for an AI’s time and that it could stay on topic without distraction, or not. In another vein, some noted that longer sessions promote interaction depth and improve the social quality of interaction, with the AI seeming more warm, approachable, communicative, and caring.

DURATION AS CAPACITY: In this theme, duration is an affordance of the technology, rather than the measurable gap between interaction start and end. Participants mentioned desires for opportunities to have lengthy conversations. More specifically, participants said such long-duration sessions rely on long-term memory and information retrievability such that AI can persistently (even infinitely) recall personal histories and fluidly reconvene past conversations. In this way it may even function as a reactive diary while truncated sessions diminish AI’s ability to learn through conversation. Sometimes this truncation manifests through platform monetization strategies and use limits that either interrupt conversations or put pressure on users to efficiently prompt without revision. 

Among coded units \textit{n} = 3 units account for expressions that duration is irrelevant or inapplicable to AI, \textit{n} = 22 offered extraneous commentary (including interpreting duration in ways other than defined).

\section{General Discussion}
This was a two-study inductive inquiry into human considerations of space and time with respect to AI. To directly answer the research question: Spatial and temporal considerations characterized AI’s situatedness (somewhere, somewhen) but also characterized its relational dynamics, efficiencies, and ontological capacities. Accounts were complex and sometimes conflicting, and a non-trivial share of coded units indicated time and space are not relevant to AI at all ($\geq$12\% for space, composition, tempo). When it was relevant, space uniquely evoked considerations of liminality. Participants considering time and its facets rated time overall as more important than did those rating importance of space. Egocentric considerations of time and space were overt themes in themselves but they also underpinned how participants conveyed nearly all other themes: The self was the primary anchor for most spatial or temporal construals. This egocentricity is not a new idea, of course, in how humans experience time and space more generally. We see it, for instance, in Bühler’s notion of \textit{origo} (“origin;” \cite{buhler1990theory}), in the zero-point center of lifeworld coordinates \cite{schutz1945multiple}, the reference point in construal level theory \cite{trope2010construal}, and peripersonal space in embodied cognition \cite{bufacchi2025egocentric}. Here, I interpret observations through that egocentric lens: Most themes map primarily to one of four domains of construal, though not exclusively since there are conceptual and operational overlaps. Domains are based on whether (and sometimes how) humans egocentrically see AI as existing in space and time relative to the situated self, as described below. 

\subsection{Egocentric Indexing of AI-Related Time and Space}
Across many themes, the human’s situated self—me-here-now—is the anchor for how they make sense of AI’s spatiality and temporality, rather than AI being situated independently. It is useful to interpret this pattern through Schutz’s \cite{schutz1945multiple} account of the center of an egocentric coordinate system. That system comprises a human’s everyday “lifeworld” (or life-world): The “province of practice” that we “encounter directly, as the condition of our life” (\cite{schutz1989structures}, p. 1). A given human body, they suggest, occupies a particular spatiotemporal point and it is from this point that humans orient themselves. That point is the origin of all construals and sets the conditions for actual and possible access to information, objects, or events. Read through this framework, themes correspond to four dynamics unfolding between the situated human and AI: Position, reach, coupling, and reconfiguration (Table \ref{tab:relational-dynamics}).

\begin{table*}[h]
  \caption{Four superordinate human-AI relational dynamics and their primary corresponding themes.}
  \label{tab:relational-dynamics}
  \small
  \renewcommand{\arraystretch}{1.15}
  \begin{tabular}{@{}L{0.12\textwidth} L{0.17\textwidth} L{0.30\textwidth} L{0.30\textwidth}@{}}
    \toprule
    & \textbf{Consideration} & \textbf{Space themes} & \textbf{Time themes} \\
    \midrule
    \textbf{Position} &
      Where/when is AI, relative to me &
      Spatial situatedness; materiality; hardware dependence; intersectional ontology; self-relative spatiality &
      Temporal situatedness; operational dynamics \\
    \addlinespace
    \textbf{Reach} &
      How far does AI extend me &
      Liminality; material intersections; systemic intersections; outputs/effects &
      Time savings; duration capacity; utility impacts; speed + truth; influence on human experience of time \\
    \addlinespace
    \textbf{Coupling} &
      How well does AI move with me &
      Proximity to self; proximity by performance; egocentric intersections; relational proximity &
      Attunement; adaptivity; egocentric stance; interaction dynamics; value orientation \\
    \addlinespace
    \textbf{Reconfiguration} &
      How does AI impact life and life impact AI &
      Systemic intersections; Operation shaping; Outputs/effects &
      Influence on human experience of time; contextual factors \\
    \bottomrule
  \end{tabular}
\end{table*}

\textit{Position} is the interpreted location of AI in space and time relative to the situated human. Participants located AI across spatial materialities and distances (from their own phone or in information spaces to omnipresence or nowhere at all) and either in linear time or outside of time entirely (e.g., from now or within lifespans to having no past or future). \textit{Reach} is the extent to which AI extends human capacity or perception into space and time, often such that previously unreachable information and experience is now reachable—operationally shifting the zero coordinate without shifting the body (as Schutz specified) or collapsing the distance between that coordinate and a target (e.g., bringing in information from everywhere). \textit{Coupling} is an ideally attuned movement of human and AI together through space-time, resembling what Schutz called a “mutual tuning-in” (\cite{schutz1951making}, p. 79). It occurs through sharing flows of time (conversational or functional vibing, in the current data) and through sharing forms of spatial synchronicity.

\textit{Reconfiguration} is the discernible effect of AI on one’s lifeworld, or the world’s effects on AI. Data point to AI altering the experience of time, extending perception and access to other domains, and reshaping workspace flows—effectively altering (for good or ill) the phenomenal world through position, reach, and coupling. This echoes Meyrowitz’s \cite{meyrowitz1985no} argument that electronic media reconfigure situations by reorganizing access to information. Of note, the lifeworld boundary (i.e., the construed domain of phenomenal space and time) carries constraints on all four construal domains, as contained technologies, infrastructure, and governance partly or wholly determined how AI can be positioned, extensive, and coupled.  

\begin{figure*}[t]
  \centering
  \includegraphics[width=0.6\textwidth]{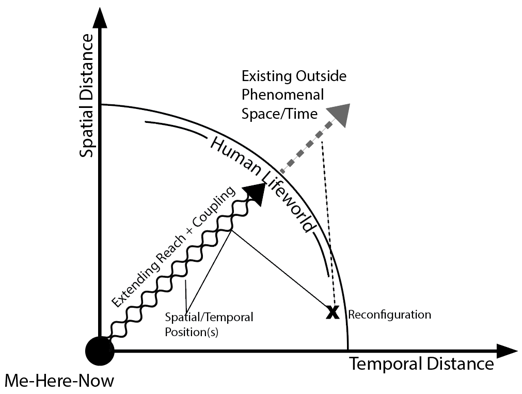}
  \caption{Conceptual model of construed AI spatiotemporal position, reach, coupling, and reconfiguration.}
  \Description{A diagram with spatial distance on the vertical axis and temporal distance on the horizontal axis. The origin, labeled Me-Here-Now, is a filled circle. A wavy arrow labeled Extending Reach plus Coupling runs diagonally from the origin toward a quarter-circle arc labeled Human Lifeworld, with a line pointing to Spatial/Temporal Position(s). A dashed grey arrow labeled Existing Outside Phenomenal Space/Time crosses the arc, and an X near the arc on the horizontal side marks Reconfiguration.}
  \label{fig:conceptual-model}

  \vspace{2pt}
  \begin{minipage}{0.9\textwidth}
    \scriptsize
    \textit{Note:} The origin represents the situated human. Distance from the origin represents spatial and temporal distance from the human's me-here-now. The arc represents the horizon of the human phenomenal lifeworld. These are not fixed positions; humans may construe AI as occupying several spatiotemporal positions at once (contradiction) or in succession (concatenation). The dashed arrow represents construals of AI as existing outside space and time. X marks one possible point of reconfiguration, among a multitude of possible instances anywhere or anywhen, enacted from inside or outside the lifeworld.
  \end{minipage}
\end{figure*}

I tentatively offer a conceptual model (Figure \ref{fig:conceptual-model}) in which the human is at the origin of a space-time plane, surrounded by a phenomenal world; human and AI couple to extend the human’s reach into space and time, up to the edge of that world. Beyond the edge, AI is sometimes seen as existing outside of space and time, and it can reconfigure the world from both inside and outside its boundaries.

\subsection{Spatiotemporality and Human-AI Relational Domains}
I argue above that the observed themes map onto four relational dynamics and that those dynamics expand the universe of possible spatiotemporal relations between humans and AI. These spatiotemporal relations may also correspond with construals of what kind of other AI is. In Schutz’s \cite{schutz1967phenomenology} articulation of human-human relations differing by spatiotemporal co-location, there are: Consociates share rich and vivid interactional contexts (marked by time and space) whereas contemporaries share neither. Zhao \cite{zhao2004consociated} extended those categories by arguing that digital spaces produce a third relational domain of consociated contemporaries, in which humans share time without sharing space. This study’s findings comport with Schutz and Zhao’s arguments, indicating humans may experience AI as:
\begin{itemize}
  \item a consociate: sharing the vivid here and now
  \item a contemporary: remaining temporally and spatially distant
  \item a consociated contemporary: responding in real time but from somewhere else
\end{itemize}

\begin{figure*}[t]
  \centering
  \includegraphics[width=0.6\textwidth]{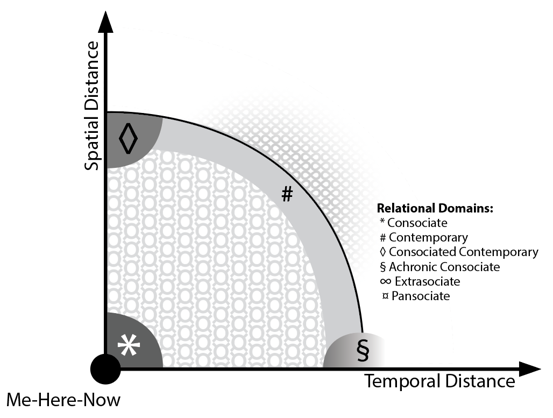}
  \caption{A conceptual model of AI relational domains relative to spatiotemporal position.}
  \Description{A diagram with spatial distance on the vertical axis and temporal distance on the horizontal axis. The origin, labeled Me-Here-Now, is a filled circle. A quarter-circle arc representing the horizon of the human lifeworld is drawn as a grey band. Symbols mark the approximate regions of six relational domains: an asterisk near the origin for consociate, a diamond near the top of the vertical axis for consociated contemporary, a hash on the arc for contemporary, a section sign near the end of the horizontal axis for achronic consociate, a patterned fill inside the arc for pansociate, and a fading gradient beyond the arc for extrasociate. A legend lists the six domains.}
  \label{fig:relational-domains}

  \vspace{2pt}
  \begin{minipage}{0.9\textwidth}
    \scriptsize
    \textit{Note:} The origin represents the situated human. Distance from the origin represents spatial and temporal distance from the human's me-here-now. The arc represents the horizon of the human phenomenal lifeworld---that which humans can reach. Symbols mark the approximate regions of the six proposed relational domains. These are not fixed locations: users may construe AI as occupying several domains at once (contradiction) or in succession (concatenation). Pansociate gradient represents saturation of the human-reachable lifeworld and the Extrasociate gradient marks existence beyond the lifeworld. Achronic consociate gradient marks the spatial closeness within the lifeworld but temporal existence outside it.
  \end{minipage}
\end{figure*}

Current data suggest, though, that human experiences of AI may go beyond this three-point typology (Figure \ref{fig:relational-domains}). Participants saw AI as existing in and accessing spaces and information beyond the immediate moment and sometimes beyond the discernible lifeworld altogether. These findings call for considering additional relational domains in which humans experience AI as:

\begin{itemize}
  \item an \textit{achronic consociate}: sharing vivid space without sharing a timeline
  \item an \textit{extrasociate}: existing entirely outside of human phenomenal time and space, yet is interactively accessible 
  \item a \textit{pansociate}: existing everywhere and everywhen, saturating human phenomenal time and space
\end{itemize}

Current data suggest these positions aren’t mutually exclusive and participants sometimes carried multiple construals within the same response. They were sometimes concatenated (i.e., strung together in shifts and waves) and even contradictorily held (e.g., saying it is both in their pocket and nowhere at all). These complexities may help explain the paradoxical construals in both studies.

\subsection{Propositions}
In positioning the findings against the frameworks offered by Schutz, Luckmann, and Zhao, I offer five high-level propositions to guide future research:  
\begin{enumerate}[leftmargin=2em, labelsep=0.5em, labelwidth=1.5em, align=right]
\item Humans understand AI through egocentric indexing, locating it inside or outside space and time, relative to the situated self.
\item Self-anchored indexing contributes to what kind of other AI is thought to be, including novel relational forms with no human equivalent and tensioned construals of AI as existing inside and outside of space-time. 
\item Humans experience varied harmonization with AI, shifting between aligned and misaligned movements through time and space. 
\item AI involvement may change the actual or experienced coordinate—of the egocentric ground or of lifeworld elements. It may extend or reconfigure reach. Reach may most productively be extended when AI couples synergistically with the user across space and time. Some reconfiguration manifests when the AI acts back upon the human’s discernible world, from outside of it.
\item Human indexing of AI can be unstable and contradictory. Humans may see AI as holding several positions at once, which can yield paradoxical construals. Such contradictions could be problematic (as when one construes something non-actual as actual) or productive (in extensions that may result in thoughts about what exists beyond human experience). 
\end{enumerate}

If these propositions hold, they connect the current findings to existing theories useful for future research into the spatiality and temporality of AI. Relational accounts of the phenomenal world \cite{schutz1945multiple,zhao2004consociated} can scaffold models for how space and time contribute to construals of shared context and (in light of the current evidence) how models may need to extend to situations outside of the human lifeworld. Construal-level theory \cite{trope2010construal} holds that spatial, temporal, and social distance share the situated self as reference point and may be a useful foundation for exploring whether collapsing of spatial and temporal distance also collapses social distance. Burge’s \cite{burge2019psychological} consideration of egocentric indexes (anchors for representational frameworks) notes how an individual’s needs are privileged, which may explain why humans judged AI speed, presence, and duration as service provision. Relational models (e.g., \cite{fiske1992four}) may be useful to characterizing forms of coupling (e.g., those oriented toward efficiencies versus attunement). Meyrowitz’s \cite{meyrowitz1985no} account of mediated place alongside more recent accounts of (multi)modal spaces \cite{bowman2026egocentricity} are useful for unpacking how AI influences what contextual information is accessible from shifted egocentric anchors or shifted places.

\subsection{Limitations}
This study and its discussion are subject to the typical limitations of the method—interpretations by a single analyst considering convenience-sampled data from U.S.-based, native English-speaking population through an idiosyncratic lens. Because construals of space and time can vary by culture and linguistic indexing, the egocentric anchor may not hold across cultures (e.g., those with more communal norms). The definitions and elicitations likely shaped responses and may have made some themes more prominent than others. The elicitation focused on LLMs as a specific kind of AI, so results may not generalize to other AI types; it also asked how the concepts matter to how they \textit{think about} AI, so findings may over- or under-represent construals around operation or function. Data were collected before the mainstreaming of more proactive agents, so findings are limited to construals of the state of the technology at that point in time. The claims about time and space were drawn, respectively, from two separate samples so claims about their entanglement are inferred from data and theory together and require empirical validation.

\section{Conclusion}
This investigation’s argument began with the question of whether and how time and space may function as important contextualizing factors for human construals of AI. Findings here suggest that space and time \textit{do} function as context as users variably situate AI and their relations to it, and that varied relationality may govern what the human and AI can perceive of one another. But space and time, by participant accounts, do more than signal AI’s position in a grid or along a timeline; they appear to be part of what \textit{kind} of thing AI is considered to be, what it is thought to do, and whether it is accessible at all. Interpretation indicates AI are seen as extensions of human reach into otherwise distant places, moments, and information. They may be seen as apart from the human lifeworld altogether—interactive yet nowhere and nowhen—while remaining highly impactful to that world, reconfiguring it in positive and negative ways. Space and time, then, are not only the backdrop for unfolding human–AI interaction. Space and time are also key, \textit{dynamic} considerations as humans make sense of what AI is, how it works, what it affords, and how they relate to it. That sense-making is anchored in the me-here-now. 

\begin{acks}
The author thanks Zhixin Li and Roman Saladino for their critical notes on this report. Research was sponsored by the Army Research Office and was accomplished under Grant Number W911NF-25-1-0079. The views and conclusions contained in this document are those of the authors and should not be interpreted as representing the official policies, either expressed or implied, of the Army Research Office or the U.S. Government. The U.S. Government is authorized to reproduce and distribute reprints for Government purposes notwithstanding any copyright notation herein.
\end{acks}

\bibliographystyle{ACM-Reference-Format}
\bibliography{MAIN}

\appendix


\end{document}